\documentclass[superscriptaddress,nofootinbib,notitlepage,twocolumn]{revtex4-2}
\pdfoutput=1
\usepackage{graphicx}
\usepackage{amsthm}
\usepackage{thmtools}
\usepackage{mathtools}
\usepackage{thm-restate}
\usepackage{amsmath}
\usepackage{amssymb}
\usepackage{bm}
\usepackage{xcolor}
\usepackage{placeins}

\usepackage[normalem]{ulem}
\usepackage[colorlinks]{hyperref}
\hypersetup{
	pdfstartview={FitH},
	pdfnewwindow=true,
	colorlinks=true,
	linkcolor=blue,
	citecolor=blue,
	filecolor=blue,
	urlcolor=blue}
\usepackage{tikz}
\usetikzlibrary{calc,backgrounds}
\usepackage{physics}
\usepackage{cancel}
\usepackage{multirow}
\usepackage{pgffor}
\usepackage{url}
\usepackage{hypcap}
\usepackage{dsfont}
\usepackage{soul}
\usepackage{inconsolata}
\usepackage{hyperref}
\usepackage{footnote}	
\usepackage{qcircuit}
\usepackage{algorithm}
\usepackage{algpseudocode}
\hypersetup{pdfpagemode=UseNone}

\allowdisplaybreaks

\newcommand{\thm}[1]{\hyperref[thm:#1]{Theorem~\ref*{thm:#1}}}
\newcommand{\defn}[1]{\hyperref[defn:#1]{Definition~\ref*{defn:#1}}}
\newcommand{\lem}[1]{\hyperref[lem:#1]{Lemma~\ref*{lem:#1}}}
\newcommand{\prop}[1]{\hyperref[prop:#1]{Proposition~\ref*{prop:#1}}}
\newcommand{\fig}[1]{\hyperref[fig:#1]{Figure~\ref*{fig:#1}}}
\newcommand{\tab}[1]{\hyperref[tab:#1]{Table~\ref*{tab:#1}}}
\renewcommand{\sec}[1]{\hyperref[sec:#1]{Section~\ref*{sec:#1}}}
\newcommand{\app}[1]{\hyperref[app:#1]{Appendix~\ref*{app:#1}}}
\newcommand{\cor}[1]{\hyperref[cor:#1]{Corollary~\ref*{cor:#1}}}
\newcommand{\obs}[1]{\hyperref[obs:#1]{Observation~\ref*{obs:#1}}}
\newcommand{\nn}{\nonumber \\}

\usepackage[capitalise]{cleveref} 

\newtheorem{theorem}{Theorem}

\renewcommand{\ket}[1]{|#1\rangle}

\newcommand{\MQ}{\affiliation{%
School of Mathematical and Physical Sciences,
Macquarie University, Sydney, NSW 2109, Australia} }
\newcommand{\UT}{\affiliation{University of Toronto, Department of Computer Science, Toronto ON, Canada}}

\begin{document}

\title{Doubling Efficiency of Hamiltonian Simulation via Generalized Quantum Signal Processing}
\author{Dominic W. Berry}
 	\email{Electronic mail: dominic.berry@mq.edu.au}\MQ

\author{Danial Motlagh}
\affiliation{Xanadu, Toronto, ON, M5G 2C8, Canada}
\UT

\author{Giacomo Pantaleoni}\MQ

\author{Nathan Wiebe}
\UT
\affiliation{Pacific Northwest National Laboratory, Richland WA, USA}
\affiliation{Canadian Institute for Advanced Research, Toronto ON, Canada}

\begin{abstract}
    Quantum signal processing provides an optimal procedure for simulating Hamiltonian evolution on a quantum computer using calls to a block encoding of the Hamiltonian.
    In many situations it is possible to control between forward and reverse steps with almost identical cost to a simple controlled operation.
    We show that it is then possible to reduce the cost of Hamiltonian simulation by a factor of 2 using the recent results of generalised quantum signal processing.
\end{abstract}

\maketitle

Early quantum algorithms for Hamiltonian simulation were based on product formulae \cite{Lloyd,BerryCMP07}, but more advanced procedures were developed based on quantum walks \cite{Childs2009,BerryQIC12} and linear combinations of unitaries \cite{LCU,BerrySTOC14}.
Those algorithms provided complexity scaling logarithmic in the inverse error $1/\epsilon$ which is optimal, but the joint scaling between $\log(1/\epsilon)$ and the time $t$ was not optimal, because it was in the form of a product rather than a sum.

Further advances using corrections improved on that $\mathcal{O}(t\log(1/\epsilon))$ complexity \cite{BerryQIC16,Novo2017}, but the true additive complexity was not achieved until the development of quantum signal processing by Low and Chuang \cite{LC17}.
That proposal was based on quantum walk steps for Hamiltonians, but Low and Chuang further generalised the procedure by using qubitisation in Ref.~\cite{Low2019}.
Qubitisation generalises the prior proposals for quantum walks and linear combinations of unitaries by using \emph{block encoding}.

In block encoding a Hamiltonian, an ancilla system is used and a unitary $V$ is chosen such that 
\begin{equation}
    (\bra{0}\otimes \openone)V(\ket{0}\otimes \openone)=\frac{H}{\lambda} \, ,
\end{equation}
where $\ket{0}$ is a state for the ancilla system.
That is, preparing $\ket{0}$ on the ancilla system, applying $V$ jointly between the ancilla and target systems, then projecting onto $\ket{0}$ for the ancilla, results in $H$ applied to the target system up to a multiplicative constant of $1/\lambda$.
The quantum walk step is then constructed by combining $V$ with a reflection about $\ket{0}$ on the ancilla system as
\begin{equation}
    U = i (2\ket{0}\bra{0}\otimes \openone_s - \openone) V ,
\end{equation}
where $\openone_s$ is the identity on just the target system, and $\openone$ is on both the ancilla and target.
For this construction $V$ should be self-inverse, and it is easy to make it so in most cases.
Each eigenvalue $E_k$ of $H$ then results in two eigenvalues
$\pm e^{\pm i \arcsin(E_k/\lambda)}$ for $U$.

Alone, this walk operator does not directly provide us with a dynamic simulation.
A technique is needed to transform the walk operator $U$ into the Hamiltonian evolution $e^{-iHt}$.
That is, both eigenvalues above need to be transformed into $e^{-iE_kt}$.
If we write the eigenvalues of $U$ as $e^{i\theta}$, then the two eigenvalues of $U$ correspond to $\theta=\arcsin(E_k/\lambda)$ and $\theta=\pi-\arcsin(E_k/\lambda)$.
In both cases it is found that $\sin\theta=E_k/\lambda$, so transforming the eigenvalues of $U$ as $e^{i\theta}\mapsto e^{-i\tau\sin\theta}$ with $\tau:=t\lambda$ gives the desired Hamiltonian evolution.

In quantum signal processing~\cite{LC17,gilyen2019quantum,motlagh2023generalized} this transformation is achieved by performing applications of $U$ controlled by an ancilla qubit.
For an eigenstate of $U$, this controlled operator corresponds to an effective $X$ rotation on the control qubit dependent on $\theta$.
By choosing a sequence of controlled operations interspersed with $Z$ rotations with carefully chosen angles, it is possible 
to enact a transformation of the form~\cite{LC17}
\begin{equation}\label{eq:LC17}
     F_\theta:=\begin{bmatrix}
        A(\theta) +iB(\theta) & iC(\theta) + D(\theta) \\ iC(\theta) - D(\theta) & A(\theta) -iB(\theta)
    \end{bmatrix}.
\end{equation}
Low and Chuang then observed that
\begin{equation}
    (\bra{+}\otimes \openone) F_\theta (\ket{+}\otimes \openone) = A(\theta) + iC(\theta),
\end{equation}
which implies that if $A(\theta) = \cos(\tau\sin(\theta))$ and $C(\theta)=-\sin(\tau\sin(\theta))$, but $B(\theta)=D(\theta)=0$, then this will transform the walk's eigenvalues to the correct eigenvalues for $e^{-iHt}$.
In the following we will ignore the minus sign on $C(\theta)$ for simplicity, because it does not affect the analysis.
In practice it isn't possible to obtain these exact equalities, but they can be obtained with high accuracy with $\mathcal{O}(\lambda t+\log(1/\epsilon))$ steps.
This number of steps yields simulations that have optimal combined scaling in both $\epsilon$ and $t$~\cite{LC17}.

Subsequent to this, it was realised that in situations requiring estimation of eigenvalues it is possible to perform phase estimation directly on the walk step, avoiding the need to perform Hamiltonian simulation \cite{BerryNPJ18,poulin2018quantum} and further that the phase returned by each application of the walk can be doubled by a simple trick.
This trick involves noting that the walk operator $U$ step can be flipped to its inverse by moving the reflection before the block encoding unitary $V$.
For such an application one can take advantage of the control between forward and reverse quantum walk steps to provide a factor of 2 improvement in complexity for eigenvalue estimation \cite{BabbushPRX18}.
That then raises the question of whether such an improvement is possible for Hamiltonian simulation.
In this work we show that it is indeed possible.

In Hamiltonian simulation via quantum signal processing as described in Ref.~\cite{LC17}, there are controlled operations of the walk operator selecting between the identity and the walk operator $U$ (Fig.~1(b) of \cite{LC17}).
The number of controlled operations is denoted $N$, but in Theorem 1 of that work the order is $N/2$.
If the eigenvalue of $U$ is $e^{i\theta}$, then the controlled operation (on the system eigenstate) has eigenvalues of $1$ and $e^{i\theta}$.
As explained below Eq.~(5) of \cite{LC17}, by alternating how the controlled operators are performed one can equivalently obtain the rotation operator $\hat R_\phi(\theta)$ (page 1 of \cite{LC17}) with eigenvalues of $e^{\pm i\theta/2}$.

We propose the following strategy to obtain the doubling observed in~\cite{BabbushPRX18} by using the above approach but changing it to use a controlled operation that selects between $U$ and $U^\dagger$.  This would then implement $\hat R_\phi(2\theta)$ with eigenvalues of $e^{\pm i\theta}$.
Then the upper limits for the sums over the values of $k$ in (iii) and (iv) in Theorem 1 of \cite{LC17} would be doubled to read
\begin{align}
    {\rm (iii)} ~&~ A(\theta) = \sum_{k~{\rm even}=0}^{N} P_K \cos(k\theta)\, , \\
    {\rm (iv)} ~&~ C(\theta) = \sum_{k~{\rm even}=2}^{N} Q_K \sin(k\theta)\, .
\end{align}
In order to approximate the Hamiltonian evolution, we aim to obtain $A(\theta)$ and $C(\theta)$ given by the Jacobi-Anger expansion (Eq.~(16) of \cite{LC17})
\begin{align}
    A(\theta) &\approx \cos[\tau\sin(\theta)] = J_0(\tau) + 2\sum_{k~{\rm even}>0}^\infty J_k(\tau) \, \cos(k \theta), \\
    C(\theta) & \approx \sin[\tau\sin(\theta)] = 2\sum_{k~{\rm odd}>0}^\infty J_k(\tau) \, \sin(k \theta).
\end{align}
The problem is now that the control between $U$ and $U^\dagger$ gives $C(\theta)$ with only \emph{even} orders, whereas for the Jacobi-Anger expansion we need $C(\theta)$ with only \emph{odd} orders.

We can address this issue by multiplying $C(\theta)$ by $e^{i\theta}$.
This function now contains only even powers of $e^{i\theta}$, but it is now complex, so the result as presented in Ref.~\cite{LC17} no longer applies.
Instead we can use the result as given in Ref.~\cite{motlagh2023generalized}.
That work provides the theorem below.
\begin{theorem}[Generalized Quantum Signal Processing]\label{thm:PolynomialOfUnitaries}
$\forall d\in \mathbb{N},\>\exists\> \vec{\theta}, \vec{\phi} \in \mathbb{R}^{d+1},\>\lambda \in \mathbb{R}$ s.t:
\[
    \left( \prod_{j=1}^{d} {R(\theta_{j}, \phi_{j}) \begin{bmatrix}
        U & 0\\        
        0 & \openone \\
    \end{bmatrix}} \right) R(\theta_0, \phi_0, \lambda) = 
    \begin{bmatrix}
        P(U) & .\>\>\>  \\        
        Q(U) & .\>\>\> \\
    \end{bmatrix}
\]
if and only if
        \begin{enumerate}
            \item  $P, \,Q\in \mathbb{C}[z]$ and $\text{deg}(P), \text{deg}(Q) \leq d$.
            \item  $\forall z\in \mathbb{C},\> |z|=1 \implies |P(z)|^2 + |Q(z)|^2 = 1$.
        \end{enumerate}
\end{theorem}

Where $R(\theta, \phi, \lambda)$ is a parameterised SU(2) rotation on the ancilla qubit. For succinctness we have written $R(\theta_{j}, \phi_{j})$ in place of $R(\theta_{j}, \phi_{j}, 0)$.
The functions $P(z)$ and $Q(z)$ are now general complex functions, as opposed to $A(\theta)$ and $C(\theta)$ being real functions above.

The convention in Ref.~\cite{motlagh2023generalized} is that only controlled $U$ operations are performed, so the polynomials $P(U)$ and $Q(U)$ contain only non-negative powers of $U$.
Now if we were to perform controlled applications of $U^2$, then we would obtain only positive even powers of $U$.
If we perform $N$ controlled operations then we obtain a maximum power $2N$ of $U$.
At the end we could consider performing $(U^\dagger)^N$, which would then give us powers of $U$ from $-N$ to $N$ in steps of 2.
But, performing $N$ controlled applications of $U^2$ followed by $(U^\dagger)^N$, is logically equivalent to performing $N$ controlled applications of $U$ versus $U^\dagger$.

We therefore reformulate the protocol above to use directionally controlled unitaries.
\begin{theorem}\label{thm:ModifiedPolynomialOfUnitaries}
$\forall d\in \mathbb{N},\>\exists\> \vec{\theta}, \vec{\phi} \in \mathbb{R}^{d+1},\>\lambda \in \mathbb{R}$ s.t:
\[
    \left( \prod_{j=1}^{d} {R(\theta_{j}, \phi_{j}) \begin{bmatrix}
        U & 0\\        
        0 & U^\dagger \\
    \end{bmatrix}} \right) R(\theta_0, \phi_0, \lambda) = 
    \begin{bmatrix}
        P(U) & -Q(U)^\dagger  \\        
        Q(U) & P(U)^\dagger \\
    \end{bmatrix}
\]
if and only if
        \begin{enumerate}
            \item  $P, \,Q\in \mathbb{C}[z^{-1}, z]$ and $\text{deg}(P), \text{deg}(Q) \leq d$.
            \item  $\text{Parity}(P), \text{Parity}(Q) = d \>\text{mod}\>2$.
            \item  $\forall z\in \mathbb{C},\> |z|=1 \implies |P(z)|^2 + |Q(z)|^2 = 1$.
        \end{enumerate}
\end{theorem}
Here, we have provided the explicit form for all 4 blocks of polynomials generated using the above procedure; the proof for this form is given in \app{block}.
Theorem~\ref{thm:ModifiedPolynomialOfUnitaries} implies that the result of Ref.~\cite{motlagh2023generalized} equally well applies to generating functions $P$ and $Q$ that have positive and negative powers of $U$.
The operator $U$ is effectively of the form $U = e^{i\mathcal{H}} = e^{i\arcsin{(H/\lambda)}}$, ignoring $\pm$ and the action on the ancilla for simplicity.
We can therefore produce
\begin{align}
    P(U) &\approx U\cos[\tau\sin(\mathcal{H})] 
    \approx \sum_{m=-K/2}^{K/2} J_{2m}(\tau) U^{2m+1} \label{eq:PU}\\
    Q(U) &\approx i\sin[\tau\sin(\mathcal{H})] \approx i \sum_{m=-K/2-1}^{K/2} J_{2m+1}(\tau) U^{2m+1} \, .
\end{align}
Here $K$ is even and $K+1=d$ is the number of controlled operations.
In this approximation we have only odd powers of $U$, but they are in steps of 2 as required.
The key improvement here is that the order is the same as the number of controlled operations, not \emph{half} as it is in standard quantum signal processing from Ref.~\cite{LC17} (where the number of controlled operations is denoted $N$).

We then remove the extra factor of $U$ in $P(U)$ by applying an initial controlled $U$ and a final controlled $U^\dagger$ to give
\begin{equation}\label{eq:correction}
\begin{bmatrix}
        U^\dagger & 0  \\        
        0 & \openone
    \end{bmatrix}
    \begin{bmatrix}
        P & -Q^\dagger  \\        
        Q & P^\dagger
    \end{bmatrix}
    \begin{bmatrix}
        \openone & 0  \\        
        0 & U
    \end{bmatrix}
    =
    \begin{bmatrix}
        U^\dagger P & -Q^\dagger \\        
        Q & P^\dagger U
    \end{bmatrix}.
\end{equation}
In this form we have the blocks corresponding to approximately $\cos[\tau\sin(\mathcal{H})]$ and $i\sin[\tau\sin(\mathcal{H})]$ similar to the form in Eq.~\eqref{eq:LC17} from Ref.~\cite{LC17}.
Therefore, using $\ket{+}$ on the ancilla qubit yields the Hamiltonian evolution on the target system.

Now the remaining problem is the requirement that $|P|^2+|Q|^2$ is \emph{exactly} equal to 1.
In the truncated form this sum will not be exactly 1 because the Jacobi-Anger expansion is necessarily truncated.
In fact, if we use the truncated sums as above then $|P|^2+|Q|^2$ can be slightly larger than 1.
That problem is avoided by simply multiplying those truncated sums by a factor slightly less than 1 to ensure that the sum is no larger than 1 (as used in Ref.~\cite{LC17}).
We then need to find new functions to add such that the sum of squares is exactly 1.

For convenience we define
\begin{align}
    P_K(U) &:= J_0(\tau) + \sum_{m=-K/2}^{K/2} J_{2m} U^{2m}\, , \\
    Q_K(U) &:= i \sum_{m=-K/2-1}^{K/2} J_{2m+1} U^{2m+1}\, ,
\end{align}
where we have removed the factor of $U$ from $P$.
If $\alpha<1$ is the factor needed to avoid the sum of squares being larger than 1,
we need to find real functions $P',Q'$ such that
\begin{align}
    &| \alpha P_K + i P' |^2 + | \alpha Q_K + Q' |^2 \nn
    &= |\alpha P_K|^2 + |\alpha Q_K|^2 + |P'|^2 + |Q'|^2 =1 \, .
\end{align}
This can alternatively be formulated as finding $P',Q'$ such that
\begin{equation}
    P'^2 + Q'^2 = 1 - \alpha^2 (P_K^2 - Q_K^2) \, .
\end{equation}
This is a problem of finding two polynomials such that the sum of squares is equal to 1.
The subtlety here is that the parities of $P_K$ and $P'$ must match, as must those of $Q_K$ and $Q'$.
This is because we will be generating $U(\alpha P_K+P')$ and $\alpha Q_K+Q'$ using the procedure of Theorem~\ref{thm:ModifiedPolynomialOfUnitaries}, and these must have $U$ to only odd orders.
This means that $P'$ must have only odd orders and $Q'$ must have only even.
Our challenge is therefore to construct $P',Q'$ with the correct parity to give the sum of squares.

To be more specific, we can rewrite $P_K,Q_K$ as
\begin{align}
    P_K(\theta) 
    &= J_0(\tau) + 2\sum_{k~{\rm even}>0}^K J_k(\tau) \, \cos(k \theta) \nn
    &= J_0(\tau) + 2\sum_{k~{\rm even}>0}^K J_k(\tau) \, T_k(\cos(\theta)) \nn
    &= J_0(\tau) + 2\sum_{m=1}^{K/2} J_{2m}(\tau) \, T_{2m}(x) \, , \\
    Q_K(\theta)
    &= 2i\sum_{k~{\rm odd}>0}^K J_k(\tau) \, \sin(k \theta) \nn
    &= 2i\sum_{k~{\rm odd}>0}^K J_k(\tau) \, \sin(\theta) \, U_{k-1}(\cos(\theta)) \nn
    &= 2\sqrt{x^2-1} \sum_{m=0}^{K/2} J_{2m+1}(\tau) \, U_{2m}(x) \, ,
\end{align}
where $x=\cos(\theta)$.
We can then construct $P',Q'$ by analysis of polynomials in $x$.

We will analyse the case where the polynomial we wish to construct $1 - \alpha^2(P_K^2 - Q_K^2)$ is non-negative.
We can then use reasoning based on Ref.~\cite{Rudin2000}.
A non-negative polynomial $p(x)$ can be written as
\begin{equation}
    p(x) = \prod_j (x-c_j)^2 \prod_k [(x-a_k)^2 + b_k^2] \, ,
\end{equation}
where $c_j,a_k,b_k$ are real numbers.
Since $p(x)$ must be non-negative for all real $x$, we will need to reason about the case where $|x|>1$ even though here we only use $|x|\le 1$.
If we first choose
\begin{equation}
    q(x) = \prod_j (x-c_j) \prod_k [x-a_k + i b_k] \, 
\end{equation}
then $p(x)$ may be expressed as a sum of squared polynomials as~\cite{Rudin2000}
\begin{equation}
    p(x) = ({\rm Re} (q(x)))^2 + ({\rm Im}(q(x)))^2 \, .
\end{equation}

In our case $p(x)$ is \emph{even}, so we know that for every root there is a corresponding negative root.
Therefore for every root-pair $a_k\pm i b_k$ there is a pair $-a_k\pm i b_k$;
that is, for every root with (non-zero) $a_k$ there is another with the sign of this real part flipped.
That implies $q(x)$ will contain pairs of factors as
\begin{equation}
    [x-a_k + i b_k] [x+a_k + i b_k] = (x^2-a_k^2-b_k^2)  + 2ib_k x \, .
\end{equation}
Similarly, for each root $c_j$ there must be a corresponding negative root, so $q(x)$ will contain pairs of factors as $(x-c_j)(x+c_j)=(x^2-c_j^2)$.
We will multiply many of these to construct $q(x)$, then at the end we will be using the real and imaginary parts for $P'$ and $Q'$.

A particularly useful feature is that in both cases the real part has even parity and the imaginary part has odd parity.
Whenever multiplying expressions where the real and imaginary parts have different parities, we obtain a resulting expression with the same property.
That is, we have the rules
\begin{align}
    (even + i\times odd) (even + i\times odd)   &\mapsto  (even + i\times odd), \nn
    (even + i\times odd) (odd + i\times even)  &\mapsto  (odd + i\times even), \nn
    (odd + i\times even)  (odd + i\times even)   &\mapsto  (even + i\times odd), \nonumber
\end{align}
where we use \emph{even} and \emph{odd} to indicate general even or odd polynomials.
What this means is that at the end $q(x)$ must have different parities for the real and imaginary parts.
Since the real and imaginary parts are used for $P',Q'$, we can obtain $P',Q'$ with the desired parities.

What remains to show is that $1 - \alpha^2(P_K^2 - Q_K^2)$ is non-negative, regardless of the value of $x$.
It is trivial that it is non-negative in the range $x\in[-1,1]$, because the Jacobi-Anger expansion gives a result close to 1 for $P_K^2 - Q_K^2$, and we choose $\alpha$ to ensure it is no larger than 1.
We will therefore only consider the case where $|x|>1$.
Our theorem can be given as follows.

\begin{theorem}\label{thm:positive}
    For $K\ge 2$ even, $0\le \tau\le K$, and $P_K,Q_K$ defined as
\begin{align}
    P_K(x) &= J_0(\tau) + 2\sum_{m=1}^{K/2} J_{2m}(\tau) \, T_{2m}(x) \, , \nn
    Q_K(x) &= 2\sqrt{x^2-1}\sum_{m=0}^{K/2} J_{2m+1}(\tau) \, U_{2m}(x) \, ,
\end{align}
the inequality $P_K^2-Q_K^2\le 1$ is satisfied for $|x|\ge 1$.
\end{theorem}

We prove this theorem in \app{proof}.
Together with the obvious result for $|x|\le 1$, this shows that the polynomial $1 - \alpha^2(P_K^2 - Q_K^2)$ is non-negative as required.

Hence, what we have shown is that it is possible to construct an approximation of Hamiltonian evolution using a number of controls between $U$ and $U^\dagger$ that is half the number of controlled unitaries normally used in quantum signal processing.
We define $P_K,Q_K$ as above and find $\alpha \approx 1$ such that $P_K^2-Q_K^2\le 1$.
Then we obtain the polynomial $1 - \alpha^2(P_K^2 - Q_K^2)\ge 0$ in $x$, and use the procedure from Ref.~\cite{Rudin2000} to find polynomials $P',Q'$ where $P'$ is even and $Q'$ is odd such that $\alpha^2(P_K^2 - Q_K^2)+P'^2+Q'^2=1$.

We then choose $P(U)=U(\alpha P_K(U) + i P'(U))$ and $Q(U)=\alpha Q_K(U) + Q'(U)$, and use the method of Theorem~\ref{thm:ModifiedPolynomialOfUnitaries} to produce these polynomials with complex coefficients.
We combine that with two extra controlled operations as in Eq.~\eqref{eq:correction} to correct the factor on $U$, and then with the ancilla qubit starting in the $\ket{+}$ state we obtain the Hamiltonian evolution on the target system.

The majority of the complexity is in using Theorem~\ref{thm:ModifiedPolynomialOfUnitaries} to produce $P(U),Q(U)$, where the number of controlled operations needed is half what it is in Ref.~\cite{LC17}.
There are two additional controlled operations used, but these are a trivial contribution to the complexity.
Hence, these modifications together enable simulation of quantum dynamics with approximately half the number of queries as in standard quantum signal processing \cite{LC17}.
For this speedup we just need to be able to control between $U$ and $U^\dagger$ with similar complexity to controlling $U$.
In most practical cases that is true, because the control can be achieved by controlling reflections on an ancilla system used in block encoding the Hamiltonian.
Therefore this approach can give a factor of 2 speedup for Hamiltonian simulation very broadly.

DWB worked on this project under a sponsored research agreement with Google Quantum AI.
DWB is also supported by Australian Research Council Discovery Projects DP210101367 and DP220101602. 
NW is supported by the US Department of Energy,
Office of Science, National Quantum Information Science Research Centers, Co-Design Center for Quantum
Advantage under contract number DE-SC0012704.

\bibliographystyle{apsrev4-1}	

\appendix

\onecolumngrid
\section{Proof of positivity of the polynomial}
\label{app:proof}

In this appendix we give the proof that the polynomial is positive, as described in Theorem \ref{thm:positive}.
\begin{proof}
For $\tau=0$, we have $P_K=1$ and $Q_K=0$, so the result is immediate. We therefore assume $0<\tau\le K$.
For $|x|\ge 1$ the two functions can be written as, using standard properties of Chebyshev polynomials,
\begin{align}
    P_K &= J_0(\tau) + 2\sum_{m=1}^{K/2} J_{2m}(\tau) \, T_{2m}(x) \nn
    &= J_0(\tau) + 2\sum_{m=1}^{K/2} J_{2m}(\tau) \, \cosh(2m\,{\rm arcosh}|x|) \nn
    &= J_0(\tau) + \sum_{m=1}^{K/2} J_{2m}(\tau) (y^{2m}+y^{-2m}) \nn
    Q_K &= 2\sqrt{x^2-1} \sum_{m=0}^{K/2} J_{2m+1}(\tau) \, U_{2m}(x) \nn
    &= 2 \sum_{m=0}^{K/2} J_{2m+1}(\tau) \, \sinh((2m+1)\,{\rm arcosh}|x|) \nn
    &= \sum_{m=0}^{K/2} J_{2m+1}(\tau) \, (y^{2m+1}-y^{-(2m+1)}) \, ,
\end{align}
where
\begin{equation}
    y = \exp({\rm arcosh}|x|) = |x|+\sqrt{x^2-1} \, .
\end{equation}
The case $|x|\ge 1$ is equivalent to $y\ge 1$.
In order to show $P_K^2-Q_K^2\le 1$, it is sufficient to show the three inequalities
\begin{align}
    P_K+Q_K &\le \exp(\tau (y-1/y)/2), \label{eq:first} \\
    P_K-Q_K &\le \exp(-\tau (y-1/y)/2), \label{eq:second} \\
    P_K+Q_K &\ge 0 \label{eq:third}.
\end{align}
Given these inequalities, multiplying the first by the second gives $P_K^2-Q_K^2\le 1$.
The third inequality is used to ensure that multiplying $P_K-Q_K$ by $P_K+Q_K$ cannot change the sign of $P_K-Q_K$.

The first inequality in Eq.~\eqref{eq:first} is obtained via noting that the Bessel functions $J_m(\tau)$ are non-negative for $m\ge \tau\ge 0$.
Because $y\ge 1$ both $y^{2m}+y^{-2m}$ and $y^{2m+1}-y^{-(2m+1)}$ are non-negative as well.
That gives us
\begin{align}
    P_K+Q_K &= \cosh(\tau (y-1/y)/2) - \sum_{m=K/2+1}^\infty J_{2m}(\tau) (y^{2m}+y^{-2m})
    \nn
    & \qquad + \sinh(\tau (y-1/y)/2) - \sum_{m=K/2+1}^{\infty} J_{2m+1}(\tau) \, (y^{2m+1}-y^{-(2m+1)}) \nn
    &\mathrel{\le} \cosh(\tau (y-1/y)/2) + \sinh(\tau (y-1/y)/2) \nn
    &= \exp(\tau (y-1/y)/2) \, .
\end{align}
We have used the fact that $2m$ and $2m+1$ are greater than $K$ for $m>K/2$, and $K\ge \tau$, so the Bessel functions are all non-negative.

For the second inequality in Eq.~\eqref{eq:second} we start by using the Maclaurin series for Bessel functions to give
\begin{align}
    \sum_{m=K/2+1}^\infty J_{2m}(\tau) \, y^{2m} - \sum_{m=K/2+1}^\infty J_{2m+1}(\tau) \, y^{2m+1} 
    &= \sum_{m=K+2}^\infty J_m(\tau) \, (-y)^{m} \nn
    &= \sum_{m=K+2}^\infty \left(-\frac{\tau y}2\right)^{m}\sum_{n=0}^\infty \frac{(-1)^n}{n!(m+n)!}\left( \frac \tau 2\right)^{2n} \nn
    &= \sum_{n=0}^\infty \frac{1}{n!}\left( \frac \tau 2\right)^{2n} z^{-n} \sum_{m=K+2}^\infty \frac 1{(m+n)!}\left(-z\right)^{m+n} \nn
    &= \sum_{n=0}^\infty \frac{1}{n!}\left( \frac \tau {2y}\right)^{n} \sum_{m=K+n+2}^\infty \frac 1{m!}\left(-z\right)^{m} \, ,
\end{align}
where $z:=\tau y/2$.
The sum over $m$ is the remainder term of order $K+n+2$ in the expansion of $\exp(-z)$.
The integral form of the remainder tells us that
\begin{align}
    \sum_{m=K+n+2}^\infty \frac 1{m!}\left(-z\right)^{m} = (-1)^n\int_0^{z} \frac {e^{-t}}{(K+n+1)!}
    \left( z-t \right)^{K+n+1} dt \, .
\end{align}
We can exchange the order of the integral and sum to give
\begin{align} \label{eq:sum2int}
    \sum_{m=K+2}^\infty J_m(\tau) \, (-y)^{m}&=\int_0^{z} dt \, e^{-t} \left( z-t \right)^{K+1} \sum_{n=0}^\infty \frac{1}{n!(K+n+1)!}\left( -\frac {\tau (z-t)} {2y}\right)^{n} \nn
    &= \int_0^{z} dt \, e^{-t} \left( z-t \right)^{K+1} J_{K+1}(\sqrt{{2\tau (z-t)}/{y}})\left( \frac {2y}{\tau (z-t)} \right)^{K/2+1/2} \nn
    &= \int_0^{z} dt \, e^{-t}  J_{K+1}(\sqrt{{2\tau (z-t)}/{y}})\left( \frac {2y(z-t)}{\tau } \right)^{K/2+1/2} \nn
    &= \frac{\tau y^{K+2}}2 \int_0^{1} dt \, e^{-t\tau y/2}  J_{K+1}(\tau\sqrt{1-t})\left( 1-t \right)^{K/2+1/2} \, .
\end{align}
The second line is obtained by using the Maclaurin series for Bessel functions again.
Now for $t\in[0,1]$ and $K\ge \tau$ we have $K+1>\tau\sqrt{1-t}$, and so the Bessel function is non-negative.

Hence the entire expression is an integral over a non-negative expression and must be non-negative.
Therefore
\begin{equation}
    \sum_{m=K/2+1}^\infty J_{2m}(\tau) \, y^{2m} \ge \sum_{m=K/2+1}^\infty J_{2m+1}(\tau) \, y^{2m+1} \, ,
\end{equation}
which implies
\begin{equation}
    \sum_{m=K/2+1}^\infty J_{2m}(\tau) \, (y^{2m}+y^{-2m}) \ge \sum_{m=K/2+1}^\infty J_{2m+1}(\tau) \, (y^{2m+1}-y^{-(2m+1)}) \, .
\end{equation}
The left and right sides are the errors in the expansions for cosh and sinh, so this implies that
\begin{align}
    \cosh(\tau (y-1/y)/2)-P_K &\ge 
    \sinh(\tau (y-1/y)/2)-Q_K \nn
    \implies \quad P_K-Q_K &\le \exp(-\tau (y-1/y)/2)\, ,
\end{align}
which gives Eq.~\eqref{eq:second} as required.

\allowdisplaybreaks[0]
Next we will show the third inequality from Eq.~\eqref{eq:third}, that $P_K+Q_K$ is non-negative.
The first positive zeros of $J_{K+1}'$ and $J_{K+1}$ are at least $K+1$ (Eq.~(10.21.3) of Ref.~\cite{NIST:DLMF}).
Both functions are positive near the origin, so $J_{K+1}'(\tau)>0$ and $J_{K+1}(\tau)>0$ for $0<\tau\le K$.
Equation~(10.6.2) of Ref.~\cite{NIST:DLMF} then gives $J_K(\tau)=J_{K+1}'(\tau)+(K+1)J_{K+1}(\tau)/\tau>J_{K+1}(\tau)$.

Pairing the positive and negative orders using $J_{-m}(\tau)=(-1)^mJ_m(\tau)$ (Eq.~(10.4.1) of Ref.~\cite{NIST:DLMF}), the expressions for $P_K$ and $Q_K$ above give
\begin{equation}\label{eq:split-top}
    P_K+Q_K=h(y)+J_{K+1}(\tau)\left(y^{K+1}-y^{-(K+1)}\right),
    \qquad
    h(y):=\sum_{m=-K}^{K}J_m(\tau)\,y^{m}.
\end{equation}
The term containing $J_{K+1}(\tau)$ is non-negative for $y\ge1$, so it is sufficient to show that $h(y)>0$ for $y\ge1$.
Applying the recurrence $mJ_m(\tau)=\frac{\tau}{2}\bigl[J_{m-1}(\tau)+J_{m+1}(\tau)\bigr]$ (Eq.~(10.6.1) of Ref.~\cite{NIST:DLMF}) term by term, shifting the summation indices, and separating the end terms gives
\begin{align}
    y\,h'(y)
    &=\sum_{m=-K}^{K}mJ_m(\tau)\,y^{m}
    =\frac{\tau}{2}\left[
      y\sum_{n=-K-1}^{K-1}J_n(\tau)\,y^{n}
      +\frac1y\sum_{n=-K+1}^{K+1}J_n(\tau)\,y^{n}\right]\nn
    &=\frac{\tau}{2}\left(y+\frac1y\right)h(y)
      -\frac{\tau}{2}\left[J_K(\tau)\left(y^{K+1}+y^{-(K+1)}\right)
      -J_{K+1}(\tau)\left(y^{K}-y^{-K}\right)\right]. \label{eq:even-derivative}
\end{align}
The last line uses $J_{-K}=J_K$ and $J_{-K-1}=-J_{K+1}$, because $K$ is even.
For $y\ge1$, the square bracket in Eq.~\eqref{eq:even-derivative} is positive because $J_K(\tau)>J_{K+1}(\tau)>0$ and $y^{K+1}+y^{-(K+1)}>y^K-y^{-K}\ge0$.
Since $h(y)\sim J_K(\tau)y^K\to+\infty$ as $y\to\infty$, $h$ attains its minimum at some $y_*\ge1$.
At this point $h'(y_*)\ge0$, since moving to the right cannot decrease $h$, even if $y_*=1$.
If $h(y_*)\le0$, the first term on the right of Eq.~\eqref{eq:even-derivative} would be non-positive and the positive square bracket would be subtracted, giving $y_*h'(y_*)<0$.
Thus $h(y)\ge h(y_*)>0$ for all $y\ge1$.
Equation~\eqref{eq:split-top} now gives $P_K+Q_K>0$, proving Eq.~\eqref{eq:third}.

Thus we have shown the three inequalities needed to give $P_K^2 - Q_K^2\le 1$, proving the theorem.
\end{proof}

\section{Block Form for Generalized Quantum Signal Processing}
\label{app:block}

Here we show that the block form of the operator implemented in Theorem~\ref{thm:ModifiedPolynomialOfUnitaries} is in fact of the form
\begin{equation}\label{eq:blockPQ}
    \begin{bmatrix}
        P & -Q^\dagger  \\        
        Q & P^\dagger \\
    \end{bmatrix} .
\end{equation}
Consider the approach for the proof of Theorem 3 of Ref.~\cite{motlagh2023generalized}. Equation (9) of that work is, for the initial rotation,
\begin{equation}
    R(\theta_0, \phi_0, \lambda) = \begin{bmatrix} e^{i(\lambda+\phi)}\cos(\theta) \openone & e^{i\phi}\sin(\theta)\openone \\
    e^{i\lambda}\sin(\theta)\openone & -\cos(\theta) \openone \\ \end{bmatrix} .
\end{equation}
Here we modify it by a global phase factor to
\begin{equation}
    R(\theta_0, \phi_0, \lambda) = \begin{bmatrix} i \, e^{i(\lambda+\phi)/2}\cos(\theta) \openone & i \, e^{i(\phi-\lambda)/2}\sin(\theta)\openone \\
    i \, e^{-i(\phi-\lambda)/2}\sin(\theta)\openone & -i \, e^{-i(\lambda+\phi)/2}\cos(\theta) \openone \\ \end{bmatrix} .
\end{equation}
This is of the form in Eq.~\eqref{eq:blockPQ} with
\begin{align}
    P = i \, e^{i(\lambda+\phi)/2}\cos(\theta) \openone , \\
    Q = i \, e^{i(\phi-\lambda)/2}\sin(\theta)\openone .
\end{align}
Next, we will show that the form in Eq.~\eqref{eq:blockPQ} holds in general by using an inductive step similar to Eq.~(10) of Ref.~\cite{motlagh2023generalized}.
First, when performing controlled operations $\ket0\bra0 \otimes U + \ket1\bra1 \otimes U^\dagger$, that step becomes
\begin{equation} 
\begin{bmatrix} e^{i\phi}\cos(\theta) U & e^{i\phi}\sin(\theta)U^\dagger \\
\sin(\theta)U & -\cos(\theta) U^\dagger \\ \end{bmatrix}
\begin{bmatrix}
\hat{P}(U) & .\>\>\>  \\
\hat{Q}(U) & .\>\>\> \\
\end{bmatrix} = \begin{bmatrix}
P(U) & .\>\>\>  \\
Q(U) & .\>\>\> \\
\end{bmatrix} .
\end{equation}
Next, we assume that the form in Eq.~\eqref{eq:blockPQ} holds for $\hat{P},\hat{Q}$, and adjust the global phase factors to give
\begin{align}
    &\begin{bmatrix} i \, e^{i\phi/2}\cos(\theta) U & i \, e^{i\phi/2}\sin(\theta)U^\dagger \\
i \, e^{-i\phi/2}\sin(\theta)U & -i \, e^{-i\phi/2}\cos(\theta) U^\dagger \\ \end{bmatrix}
\begin{bmatrix}
\hat{P} & -\hat{Q}^\dagger  \\
\hat{Q} & \hat{P}^\dagger \\
\end{bmatrix}\nonumber\\
 &= \begin{bmatrix}
i \, e^{i\phi/2}\cos(\theta) U\hat{P} + i \, e^{i\phi/2}\sin(\theta)U^\dagger\hat{Q} & -i \, e^{i\phi/2}\cos(\theta) U\hat{Q}^\dagger + i \, e^{i\phi/2}\sin(\theta)U^\dagger\hat{P}^\dagger  \\
i \, e^{-i\phi/2}\sin(\theta)U\hat{P} - i \, e^{-i\phi/2}\cos(\theta) U^\dagger\hat{Q} & -i \, e^{-i\phi/2}\sin(\theta)U\hat{Q}^\dagger - i \, e^{-i\phi/2}\cos(\theta) U^\dagger\hat{P}^\dagger \\
\end{bmatrix} .
\end{align}
If we set
\begin{align}
    P &= i \, e^{i\phi/2}\cos(\theta) U\hat{P} + i \, e^{i\phi/2}\sin(\theta)U^\dagger\hat{Q}\, , \\
    Q &= i \, e^{-i\phi/2}\sin(\theta)U\hat{P} - i \, e^{-i\phi/2}\cos(\theta) U^\dagger\hat{Q} ,
\end{align}
then we obtain
\begin{equation}
    \begin{bmatrix} i \, e^{i\phi/2}\cos(\theta) U & i \, e^{i\phi/2}\sin(\theta)U^\dagger \\
i \, e^{-i\phi/2}\sin(\theta)U & -i \, e^{-i\phi/2}\cos(\theta) U^\dagger \\ \end{bmatrix}
\begin{bmatrix}
\hat{P} & -\hat{Q}^\dagger  \\
\hat{Q} & \hat{P}^\dagger \\
\end{bmatrix} =\begin{bmatrix}
P & -Q^\dagger  \\
Q & P^\dagger \\
\end{bmatrix} .
\end{equation}
Hence we have shown the inductive step for Eq.~\eqref{eq:blockPQ}, so it must hold in general.

\end{document}